\documentclass[conference]{IEEEtran}
\usepackage[T1]{fontenc}
\usepackage[utf8]{inputenc}
\usepackage{amsmath}
\usepackage{amsthm}
\usepackage{amssymb}
\usepackage{graphicx}
\usepackage[bookmarks=true,bookmarksnumbered=true,bookmarksopen=true,bookmarksopenlevel=1,
 breaklinks=false,pdfborder={0 0 0},pdfborderstyle={},backref=false,colorlinks=false]
 {hyperref}
\hypersetup{pdftitle={Your Title},
 pdfauthor={Your Name},
 pdfpagelayout=OneColumn, pdfnewwindow=true, pdfstartview=XYZ, plainpages=false}

\makeatletter
\theoremstyle{plain}
\newtheorem{thm}{\protect\theoremname}
\theoremstyle{definition}
\newtheorem{defn}[thm]{\protect\definitionname}
\theoremstyle{plain}
\newtheorem{prop}[thm]{\protect\propositionname}
\theoremstyle{plain}
\newtheorem{lem}[thm]{\protect\lemmaname}

\usepackage[caption=false,font=footnotesize]{subfig}

\makeatother

\providecommand{\definitionname}{Definition}
\providecommand{\lemmaname}{Lemma}
\providecommand{\propositionname}{Proposition}
\providecommand{\theoremname}{Theorem}

\begin{document}
\title{Information Theory for expectation measures}
\author{\IEEEauthorblockN{Peter~Harremoës}\IEEEauthorblockA{Niels Brock\\
Copenhagen Business College\\
Copenhagen\\
Denmark\\
Email: harremoes@ieee.org}}
\maketitle
\begin{abstract}
Shannon based his information theory on the notion of probability
measures as it we developed by Kolmogorov. In this paper we study
some fundamental problems in information theory based on expectation
measures. In the theory of expectation measures it is natural to study
data sets where no randomness is present and it is also natural to
study information theory for point processes as well as sampling where
the sample size is not fixed. Expectation measures in combination
with Kraft's Inequality can be used to clarify in which cases probability
measures can be used to quantify randomness. 
\end{abstract}

\IEEEpeerreviewmaketitle{}

\section{Introduction}

In 1957 E.T. Jaynes introduced the Principle of Maximum Entropy as
a principle for doing inference under uncertainty \cite{Jaynes1957a,Jaynes1957b}.
This principle has been very influential and it has found many applications.
For instance, a die has six sides and the maximum entropy distribution
on the six sides is the uniform distribution. When entropy is defined
as a measure of uncertainty of a probability measure, one may use
the Principle of Maximum Entropy to single out a unique probability
measure, but this approach does not explain why uncertainty should
be quantified in terms of probability measures. 

On a finite alphabet maximizing entropy of $P$ under the constraint
$P\in C$ is equivalent to minimizing the KL-divergence $D\left(P\Vert U\right)$
where $U$ denotes the uniform distribution. Kullback introduced what
he called the Principle of Minimum Information Discrimination \cite{Kullback1959}.
The Principle of Maximum Entropy and the Principle of Minimum Information
Discrimination were both studied by F. Topsøe as optimization problems
\cite{Topsoe1979}. The Principle of Minimum Information Discrimination
can be justified by the Conditional Limit Theorem \cite{Csiszar1984},
so the role of the Principle of Maximum Entropy may be reduced to
a justification of using the uniform distribution as prior distribution.

Later J. Rissanen developed the Minimum Description Length principle
\cite{Rissanen1978,Grunwald2007}. In Rissanen's approach there were
only data and descriptions. The probability distributions just become
substitutes for codes via Kraft's Inequality. 

Recently, a theory of expectation measures was introduced as an alternative
to the usual Kolmogorov style of probability theory \cite{Harremoes2025a}.
This new approach allow us to distinguish between different applications
of measure theory when it is used to model randomness or uncertainty.
Some of these measures have total mass 1 and some have finite total
mass greater than 1 or less than 1. Sometimes they may even have infinite
total mass. We will both discuss a simple setup with finite texts
and a more abstract setup with point processes.

\section{Information divergence}

Divergence measures play a major role in information theory in general,
and for expectation theory we need to define it for more general measures
than $\sigma$-finite measure. 
\begin{defn}
Let $\mu$ and $\nu$ denote discrete measures on $\mathbb{A}$. Then
\emph{information divergence} is defined by 
\[
D\left(\mu\Vert\nu\right)=\sum_{a\in\mathbb{A}}f\left(\frac{\mu\left(a\right)}{\nu\left(a\right)}\right)\cdot\nu\left(a\right)
\]
where $f\left(x\right)=x\cdot\ln\left(x\right)-\left(x-1\right).$
In this definition, we will use the conventions that $f\left(\frac{0}{0}\right)\cdot0=f\left(\frac{\infty}{\infty}\right)\cdot\infty=0$
and that $f\left(\infty\right)\cdot0=\infty.$
\end{defn}
If $\mu$ and $\nu$ are probability measures then information divergence
is called \emph{Kullback divergence} or KL-divergence and it has an
interpretation as redundancy. For more general measures we will use
the following definition.
\begin{defn}
Let $\mu$ and $\nu$ denote measures on the measurable space $\left(\mathbb{A},\mathcal{G}\right)$.
Then the \emph{information divergence} is defined by
\[
D\left(\mu\Vert\nu\right)=\sup_{\mathcal{F}}D\left(\mu_{\mid\mathcal{F}}\left\Vert \nu_{\mid\mathcal{F}}\right.\right)
\]
where the supremum is taken over all finite sub-$\sigma$-algebras
$\mathcal{F}\subseteq\mathcal{G}$. 
\end{defn}
According to Gibbs' inequality $D\left(\mu\Vert\nu\right)\geq0$ with
equality if and only if $\mu=\nu.$
\begin{prop}
\label{prop:Decomp}Let $\mu$ and $\nu$ denote measures on $\left(\mathbb{A},\mathcal{G}\right),$
and assume that $D\left(\mu\Vert\nu\right)<\infty.$ Then there exists
$B\in\mathcal{G}$ such that $\mu\left(\cdot\cap B\right)$ and $\nu\left(\cdot\cap B\right)$
are $\sigma$-finite and such that $\mu\left(\cdot\cap\complement B\right)=\nu\left(\cdot\cap\complement B\right).$
In particular a Radon-Nikodym derivative $\rho=\frac{\mathrm{d}\mu}{\mathrm{d}\nu}$
exists and 
\[
D\left(\mu\Vert\nu\right)=\int\left(\rho\cdot\ln\left(\rho\right)-\left(\rho-1\right)\right)\,\mathrm{d}\nu.
\]
\end{prop}
\begin{IEEEproof}
Let $\mathcal{F}_{1}\subseteq\mathcal{F}_{2}\subseteq\dots\subseteq G$
denote a sequence of sub-algebras such that $D\left(\mu_{\mid\mathcal{F}_{j}}\left\Vert \nu_{\mid\mathcal{F}_{j}}\right.\right)\to D\left(\mu\Vert\nu\right)$
for $j\to\infty.$

Let $A_{1}^{j},A_{2}^{j},A_{3}^{j},\dots$ denote the atoms of $\mathcal{F}_{j}$.
If $\mu\left(A_{i_{k}}\right)=\nu\left(A_{i_{k}}\right)=\infty,k=1,2$
then $A_{i_{1}}$and $A_{i_{2}}$ can be merged together without changing
the value of $D\left(\mu_{\mid\mathcal{F}}\left\Vert \nu_{\mid\mathcal{F}}\right.\right).$
Without loss of generality, we may assume $A_{1}^{j}$ is the only
atom in $\mathcal{F}_{j}$ with $\mu\left(A_{i}^{j}\right)=\nu\left(A_{i}^{j}\right)=\infty.$
Since $D\left(\mu_{\mid\mathcal{F}}\left\Vert \nu_{\mid\mathcal{F}}\right.\right)<\infty$
we must have $\mu\left(A_{i}^{j}\right)<\infty$ and $\nu\left(A_{i}^{j}\right)<\infty$
for $i>1.$

Let $\mathcal{H}_{j}$ denote the $\sigma$-algebra generated by $A_{1}^{j}$
and all $\mathcal{G}$-measurable subsets of $\complement A_{1}^{j}.$
Then $\mathcal{F}_{j}\subseteq\mathcal{H}_{j}$ and 
\[
D\left(\mu_{\mid\mathcal{H}_{j}}\left\Vert \nu_{\mid\mathcal{H}_{j}}\right.\right)\to D\left(\mu\Vert\nu\right)
\]
for $j\to\infty.$ In particular 
$D\left(\mu\left(\cdot\cap A_{1}^{j}\right)\Vert\nu\left(\cdot\cap A_{1}^{j}\right)\right)\to0$
for $j\to\infty.$ Let $B=\complement\left(\bigcap_{j}A_{1}^{j}\right).$
Then 
\[D\left(\mu\left(\cdot\cap\complement B\right)\Vert\nu\left(\cdot\cap\complement B\right)\right)=0\]
implying that $\mu\left(\cdot\cap\complement B\right)=\nu\left(\cdot\cap\complement B\right).$ 

On $\complement B$ the Radon-Nikodym simply equals 1. On $B$ the
Radon-Nikodym derivative exists because the measures are $\sigma$-finite
on $B$. For the existence of a Radon-Nikodym on a $\sigma$-finite
set one may refer to any stanard textbook on measre theory, but we
may also argue more directly. It is sufficient to prove that a derivative
exists on any of the sets $A_{i}^{j},i>1$. Without loss of generality
assume that $\nu\left(A_{i}^{j}\right)=1$. Then $\frac{\mathrm{d}\mu_{\mid\mathcal{F}_{j}}}{\mathrm{d}\nu_{\mid\mathcal{F}_{j}}}$
is a martingale. Since the martingale is log-bounded it converges
pointwise $\nu$-almost surely and in norm to a function $\rho,$
which must be a Radon-Nikodym derivative of $\mu$ with respect to
$\nu$ \cite{Harremoes2005b,Harremoes2008b}. 

With a Radon-Nikodym at our disposal it is easy to prove that information
divergence is given by an integral.
\end{IEEEproof}
The above exposition is more general than what is presented in the
recent paper \cite{Leskelae2024}, that is limited to $\sigma$-finite
measure. With minor changes our exposition even works if the measures
$\mu$ and $\nu$ are replaced by \emph{valuations} i.e. set functions
defined on distributive lattices rather than $\sigma$-algebras. The
relevance of working with valuations is discussed in \cite[Sec. 3.1-3.3]{Harremoes2025a}. 

\section{Coding for empirical measures\protect\label{sec:UniversalCoding}}

Here, we will discuss a situation where empirical measures, coding
measures, maximum entropy measures and probability measures each has
its own distinct ontological status. We will define these different
types of measures and illustrate in detail how they are related in
a simple situation. The results can be generalized to more general
convex sets than the ones that will discuss here, but we will not
discuss such generalizations here, because they may obscure the ontological
distinctions between the different types of measures. 

\subsection{Coding a single text}

An \emph{empirical measure} is a measure $\mu$ such that $\mu\left(A\right)$
equals an integer or equals $\infty.$ Thus, a table of frequences
can be represented by an empirical measure. 

First we consider a situation where a single text is given, and the
text consists of a finite number of letters from the alphabet $\mathbb{A}$.
Assume that the statistics of letters in the text is given by a table
of frequencies of the letters. We will represent such a table by the
\emph{empirical measure} $\mu.$ We want to minimize the toal code
length
\[
\sum_{a\in\mathbb{A}}\ell\left(a\right)\cdot\mu\left(a\right),
\]
where $\ell:\mathbb{A}\to\left[0,\infty\right]$ is a function that
satisfies Kraft's Inequality $\sum_{a}\exp\left(-\ell\left(a\right)\right)\leq1$
\cite{Cover1991,Harremoes2013}. Equivalently, we want to minimize
\[
\sum_{a\in\mathbb{A}}\ln\left(1/Q\left(a\right)\right)\cdot\mu\left(a\right)
\]
where $Q$ runs over different \emph{coding measures}, i.e. $Q$ is
assumed to satisfy $\left\Vert Q\right\Vert \leq1,$ where $\left\Vert Q\right\Vert =\sum_{a}Q\left(a\right).$ 
\begin{defn}
Let $\mu$ be a measure on $\mathbb{A}.$ Then the \emph{entropy}
of $\mu$ can be defined as 
\[
H\left(\mu\right)=\mu\left(\mathbb{A}\right)\cdot H\left(\mu/\left\Vert \mu\right\Vert \right).
\]
\end{defn}
Then 
\begin{align*}
\sum_{a\in\mathbb{A}}\ln\left(\frac{1}{Q\left(a\right)}\right)\cdot\mu\left(a\right) & =H\left(\mu\right)+D\left(\mu\left\Vert \left\Vert \mu\right\Vert \cdot Q\right.\right)\geq H\left(\mu\right)
\end{align*}
with equality if and only if $Q=\mu/\left\Vert \mu\right\Vert .$

\subsection{Coding one out of several texts}

Next, we consider a situation where a finite set $\Omega$ of finite
texts are given. For each text $\omega\in\Omega$ the statistics of
the letters is represented by an empirical measure that will be denoted
$\mu_{\omega}$. Our goal is to code the text in such a way that the
total code length is as short as possible. Thus, we should minimize
\[
\max_{\omega\in\Omega}\sum_{a\in\mathbb{A}}\ell\left(a\right)\cdot\mu_{\omega}\left(a\right)
\]
for functions $\ell$ such that $\sum_{a}\exp\left(-\ell\right)\leq1.$
The maximum equals
\[
\max_{\mu\in C}\sum_{a\in\mathbb{A}}\ell\left(a\right)\cdot\mu\left(a\right),
\]
where $C$ denotes the convex hull of the empirical measures. We note
that the set $C$ is convex and compact. The set of functions $\ell$
that satisfies Kraft's Inequality is convex and closed. 
\begin{thm}
Let $C$ denote the convex hull of finitely may empirical measure
on a finite alphabet $\mathbb{A}.$ Let $\mathcal{L}$ denote the
set of functions $\mathbb{A}\to\text{\ensuremath{\mathbb{R}}}$ that
satisfy Kraft's Inequality. Then there exists a measure $\mu^{*}\in C$
and a function $\ell^{*}\in\mathcal{L}$ and a constant $c$ such
that 
\begin{equation}
\sum_{a\in\mathbb{A}}\ell^{*}\left(a\right)\cdot\mu\left(a\right)\leq c\label{eq:ligevaegt1}
\end{equation}
for all $\mu\in C$, and 
\begin{equation}
\sum_{a\in\mathbb{A}}\ell\left(a\right)\cdot\mu^{*}\left(a\right)\geq c\label{eq:ligevaegt2}
\end{equation}
 for all $\ell\in\mathcal{L}.$ 
\end{thm}
\begin{IEEEproof}
The set $C$ is convex and compact, and the set $\mathcal{L}$ is
convex and closed. The function 
\[
\left(\ell,\mu\right)\to\sum_{a\in\mathbb{A}}\ell^{*}\left(a\right)\cdot\mu\left(a\right)
\]
is linear in each variable and continuous. The result follows by Sion's
Minimax Theorem \cite{Sion1958}.
\end{IEEEproof}
\begin{defn}
Let $\left(\Omega,\mathcal{F},P\right)$ denote a probability space
and let $\left(\mathbb{A},\mathcal{G}\right)$ denote a measurable
space. A kernel $\omega\to\mu_{\omega}$ from $\Omega$ to $\mathbb{A}$
is called a \emph{point process} if $\mu_{\omega}$ is an empirical
measure for all $\omega\in\Omega$, and each of the measures $\mu_{\omega}$
is called an \emph{instance of the point process}. The mixture $\int\mu_{\omega}\,\mathrm{d}P\omega$
is called the \emph{expectation} measure of the point process.
\end{defn}
Since $\mu^{*}\in C$ there exists a probability measure $P=\left(p_{\omega}\right)_{\omega\in\Omega}$
such that $\mu^{*}=\sum_{\omega}p_{\omega}\cdot\mu_{\omega}.$ In
general, the probability measure $P$ is not unique. A probability
measure $P$ has an interpretation of a point process with $\mu^{*}$
as expectation measure. 
\begin{thm}
Let $C$ denote the convex hull of finitely may empirical measure
on a finite alphabet $\mathbb{A}.$ Let $\mathcal{L}$ denote the
set of functions $\mathbb{A}\to\text{\ensuremath{\mathbb{R}}}$ that
satisfy Kraft's Inequality. Assume that $\mu^{*}\in C$ and $\ell^{*}\in\mathcal{L}$
satisfy \eqref{eq:ligevaegt1} and \eqref{eq:ligevaegt2} for some
constant $c.$ Assume that $\mu^{*}=\sum_{\omega}p_{\omega}\cdot\mu_{\omega}.$
Then 
\[
\sum_{a\in\mathbb{A}}\ell^{*}\left(a\right)\cdot\mu_{\omega}\left(a\right)=c
\]
for all $i$ for which $p_{i}>0.$ Further, $\mu^{*}$ has maximal
entropy and $\ell^{*}\left(a\right)=\ln\left(\frac{\left\Vert \mu^{*}\right\Vert }{\mu^{*}\left(a\right)}\right).$ 
\end{thm}
\begin{IEEEproof}
We have 
\begin{multline*}
c=\sum_{a\in\mathbb{A}}\ell^{*}\left(a\right)\cdot\mu^{*}\left(a\right)=\sum_{a\in\mathbb{A}}\left(\ell^{*}\left(a\right)\sum_{\omega\in\Omega}p_{\omega}\cdot\mu_{\omega}\left(a\right)\right)\\
=\sum_{\omega\in\Omega}p_{\omega}\cdot\sum_{a\in\mathbb{A}}\ell^{*}\left(a\right)\cdot\mu_{\omega}\left(a\right)\leq\sum_{\omega\in\Omega}p_{\omega}\cdot c=c.
\end{multline*}
Therefore, 
\[
\sum_{a\in\mathbb{A}}\ell^{*}\left(a\right)\cdot\mu_{\omega}\left(a\right)=c
\]
for all $\omega$ for which $p_{\omega}>0.$

The minimum of $\sum_{a}\mu^{*}\left(a\right)\cdot\ell\left(a\right)$
over functions $\ell$ that satisfy Kraft's Inequality is given by
\[
\ell\left(a\right)=\ln\left(\left\Vert \mu^{*}\right\Vert /\mu^{*}\left(a\right)\right),
\]
and the minimum is $H\left(\mu^{*}\right).$ Hence, 
\[
\ell^{*}\left(a\right)=\ln\left(\left\Vert \mu^{*}\right\Vert /\mu^{*}\left(a\right)\right)\]
and $c=H\left(\mu^{*}\right).$ For any measure $\mu\in C$ we have
\begin{align*}
c & \geq\sum_{a\in\mathbb{A}}\ell^{*}\left(a\right)\cdot\mu\left(a\right)\geq H\left(\mu\right).
\end{align*}
Hence $H\left(\mu^{*}\right)=\max_{\mu\in C}H\left(\mu\right),$ i.e.
the expectation measure $\mu^{*}$ is the maximum entropy measure
in $C$. 
\end{IEEEproof}
\begin{thm}
Let $C$ denote the convex hull of a finite set of empirical measures.
Then both the optimal length function $\ell^{*}$ and the maximum
entropy measure $\mu^{*}$ on the convex set $C$ are uniquely determined.
\end{thm}
\begin{IEEEproof}
Assume that \eqref{eq:ligevaegt1} is satisfied for both $\ell_{1}^{*}$
and $\ell_{2}^{*}.$ Then 
\[
\tilde{\ell}=\frac{\ell_{1}^{*}+\ell_{2}^{*}}{2}+\ln\left(\sum_{a\in\mathbb{A}}\exp\left(-\frac{\ell_{1}^{*}\left(a\right)+\ell_{2}^{*}\left(a\right)}{2}\right)\right)
\]
is a code length function that satisfies 
\[
\sum_{a\in\mathbb{A}}\mu\left(a\right)\cdot\tilde{\ell}\left(a\right)\leq c+\ln\left(\sum_{a\in\mathbb{A}}\exp\left(-\frac{\ell_{1}^{*}\left(a\right)+\ell_{2}^{*}\left(a\right)}{2}\right)\right)
\]
for all $\mu\in C.$ This holds in particular for $\mu=\mu^{*}$,
which implies that 
\[
\ln\left(\sum_{a\in\mathbb{A}}\exp\left(-\frac{\ell_{1}^{*}\left(a\right)+\ell_{2}^{*}\left(a\right)}{2}\right)\right)=0
\]
and $\ell_{1}^{*}\left(a\right)=\ell_{2}^{*}\left(a\right).$ Thus,
$\ell^{*}$ is uniquely determined. Hence, $\tilde{\mu}^{*}$ is uniquely
determined and $\mu^{*}$ is uniquely determined.
\end{IEEEproof}
\begin{figure}[tbh]
\begin{centering}
\includegraphics[scale=0.4]{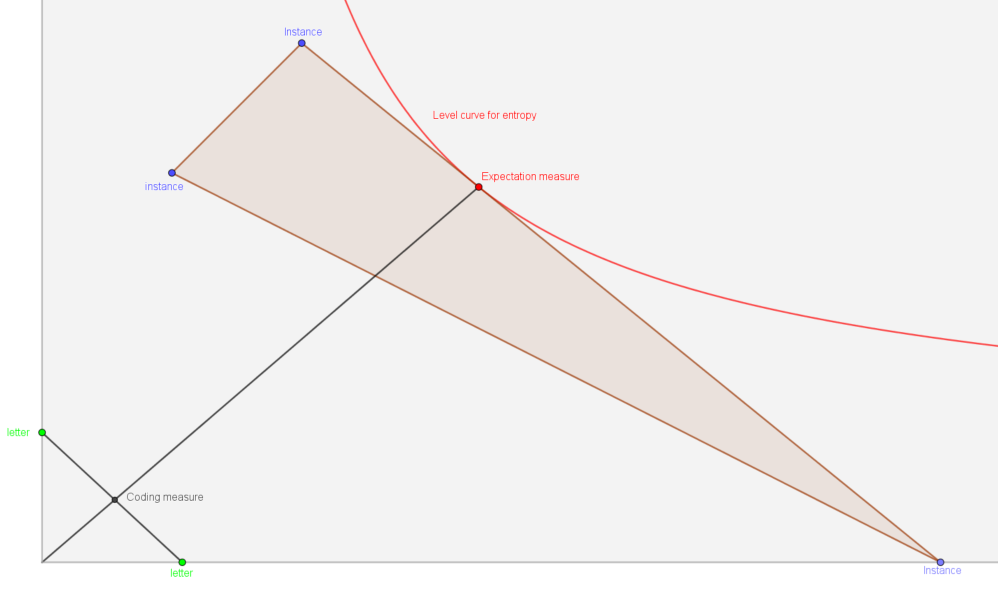}
\par\end{centering}
\caption{At the maximum entropy measure the level curve of the entropy function
is tangent to that face that contain the maximum entropy measure.}
\end{figure}

\begin{figure}[tbh]
\begin{centering}
\includegraphics[scale=0.4]{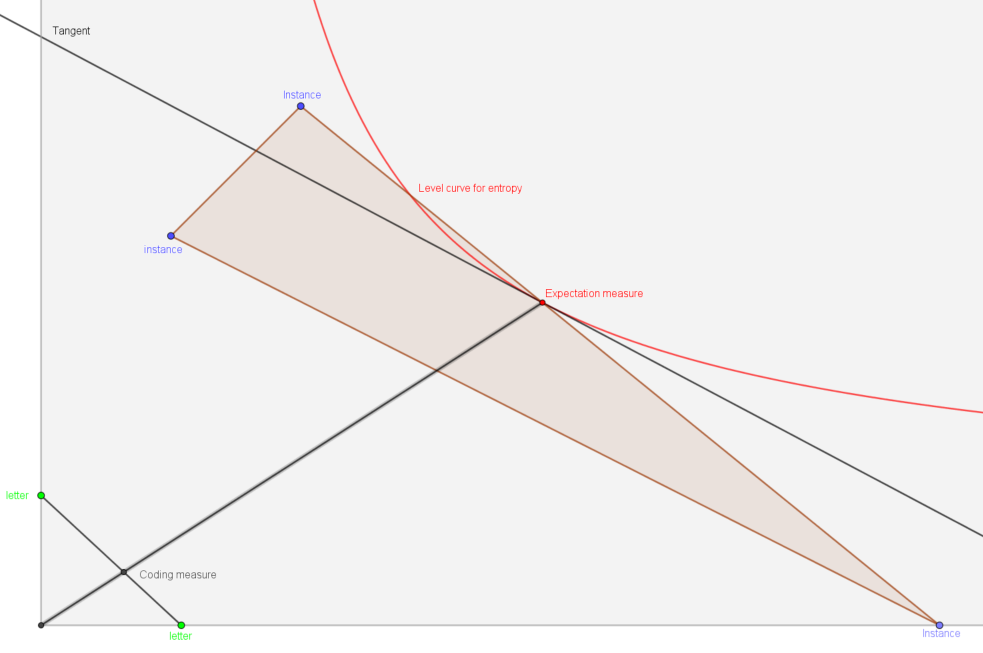}
\par\end{centering}
\caption{At a measure that does not have maximum entropy the level curve of
the entropy function is not tangent to that face that contain the
maximum entropy measure.}
\end{figure}

If a coding is based on an expectation measure in $C$ that does not
have maximum entropy then it will give higher total code length than
$c$ on some instances and lower code length than $c$ on other instances.
\begin{thm}
Let $P=\left(p_{\omega}\right)_{\omega\in\Omega}$ denote a probability
measure over the empirical measures $\mu_{\omega},\omega\in\Omega.$
Then
\[
\min_{\ell}\max_{\omega}\sum_{a\in\mathbb{A}}\ell\left(a\right)\cdot\mu_{\omega}\left(a\right)=\min_{\ell}\sum_{\omega\in\Omega}p_{\omega}\cdot\left(\sum_{a\in\mathbb{A}}\ell\left(a\right)\cdot\mu_{\omega}\left(a\right)\right)
\]
if and only $\sum_{\omega}p_{\omega}\cdot\mu_{\omega}=\mu^{*},$ where
$\mu^{*}$ is the maximum entropy measure in the convex hull of the
empirical measures.
\end{thm}
The theorem states that minimizing the \emph{maximal code length}
is not different from minimizing \emph{mean code length} with respect
to the mixing measure $P=\left(p_{\omega}\right)_{\omega\in\Omega}.$
\begin{IEEEproof}
We have
\begin{multline*}
\sum_{\omega\in\Omega}p_{\omega}\cdot\left(\sum_{a\in\mathbb{A}}\ln\left(\frac{1}{Q\left(a\right)}\right)\cdot\mu_{\omega}\left(a\right)\right)\\
=\sum_{a\in\mathbb{A}}\ln\left(\frac{1}{Q\left(a\right)}\right)\cdot\mu^{*}\left(a\right)\\
=H\left(\mu^{*}\right)+D\left(\left.\mu^{*}\right\Vert \left\Vert \mu^{*}\right\Vert \cdot Q\right)\geq H\left(\mu^{*}\right)
\end{multline*}
with equality if and only if $Q=\mu^{*}/\left\Vert \mu^{*}\right\Vert .$ 
\end{IEEEproof}
The set of instances $\mu_{i}$ for which $p_{i}>0$ is the support
for the point process $P$, and we say that the code length function
$\ell^{*}$ is \emph{cost stable} on the support of the process $P.$
\begin{thm}
Assume that $\mu^{*}$ is the maximum entropy distribution in the
convex set $C$ and assume that the corresponding code is cost stable
on $C$. Let $K\subseteq C$ be a convex subset. Then maximum entropy
measure $\mu^{**}$ on $K$ is the same as the measure $\nu\in K$
that minimizes
\[
D\left(\nu\left\Vert \mu^{*}\cdot\frac{\left\Vert \nu\right\Vert }{\left\Vert \mu^{*}\right\Vert }\right.\right).
\]
\end{thm}
\begin{IEEEproof}
The entropy of $\nu\in C$ is given by
\begin{multline*}
H\left(\nu\right)=\left\Vert \nu\right\Vert H\left(\frac{\nu}{\left\Vert \nu\right\Vert }\right)=-\sum_{a\in\mathbb{A}}\ln\left(\frac{\nu\left(a\right)}{\left\Vert \nu\right\Vert }\right)\cdot\nu\left(a\right)\\
=-\sum_{a\in\mathbb{A}}\ln\left(\frac{\mu^{*}\left(a\right)}{\left\Vert \mu^{*}\right\Vert }\right)\cdot\nu\left(a\right)-\sum_{a\in\mathbb{A}}\ln\left(\frac{\nu\left(a\right)/\left\Vert \nu\right\Vert }{\mu^{*}\left(a\right)/\left\Vert \mu^{*}\right\Vert }\right)\cdot\nu\left(a\right)\\
=-\sum_{a\in\mathbb{A}}\ln\left(\frac{\mu^{*}\left(a\right)}{\left\Vert \mu^{*}\right\Vert }\right)\cdot\mu^{*}\left(a\right)-D\left(\nu\left\Vert \mu^{*}\cdot\frac{\left\Vert \nu\right\Vert }{\left\Vert \mu^{*}\right\Vert }\right.\right)\\
=H\left(\mu^{*}\right)-D\left(\nu\left\Vert \mu^{*}\cdot\frac{\left\Vert \nu\right\Vert }{\left\Vert \mu^{*}\right\Vert }\right.\right).
\end{multline*}
Thus, minimizing $D\left(\nu\left\Vert \mu^{*}\cdot\left\Vert \nu\right\Vert /\left\Vert \mu^{*}\right\Vert \right.\right)$ is the same as maximizing $H\left(\nu\right).$
\end{IEEEproof}

\subsection{Scoring rules}

Here, we shall see how that the well-known logarithmic scoring rule
can be justified in our setup. You want to code a text $\omega$ from
a set $\Omega,$ but you do not know the elements of $\Omega$. Instead
of knowing the elements of $\Omega$ you know an expert that through
many years of study and research knows the elements of $\Omega.$
In order to code a text from $\Omega$ you just need the optimal coding
measure $\mu^{*}/\left\Vert \mu^{*}\right\Vert $ rather than precise
shape of $\Omega.$ Therefore you ask the expert for revealing the
optimal coding measure $\mu^{*}/\left\Vert \mu^{*}\right\Vert $ to
you. The expert can calculate the optimal coding measure $\mu^{*}/\left\Vert \mu^{*}\right\Vert $,
but maybe the expert is not honest and she will reveal the coding
measure $Q$ to you rather than the optimal coding measure $\mu^{*}/\left\Vert \mu^{*}\right\Vert $.
Therefore you intend to pay the expert for her service according to
the quality of the coding measure $Q$ that she reveals. This can
be done as follows. 

In a social game two agents will have an incentive to share information
when they have a common interest, i.e. when their payoff functions
are linearly dependent. Therefore, you sign a contract with the expert
so that the expert will get a fixed amount of money $f$ minus a positive
constant $k$ times 
\[
\sum_{a\in\mathbb{A}}\ell\left(a\right)\cdot\mu_{\omega}\left(a\right)
\]
where $\ell\left(a\right)=-\ln\left(Q\left(a\right)\right)$ and $\mu_{i}$
is the empirical measure for the text you end up coding. 

If the expert reveals $Q=\tilde{\mu^{*}}$ then the expert will get
a payoff of at least $f-k\cdot c$ and this payoff will be achieved
for any $\mu_{i}$ in the support of the point process $P.$ If the
expert is a Bayesian statistician and use $P$ as a prior then $f-k\cdot c$
will be the expected payoff if $Q=\tilde{\mu}^{*}$ is revealed. If
the expert reveals any other measure $Q$ then the expert will get
less expected payoff.

In conclusion, probability measures designed for minimax optimal coding
may be used by a Bayesian statistician for calculating mean payoff. 

\section{Information divergence for expectation measures}

\subsection{The Poisson interpretation}

In Section \eqref{sec:UniversalCoding} the divergence of the form
\[D\left(\nu\left\Vert \mu^{*}\cdot\left\Vert \nu\right\Vert /\left\Vert \mu^{*}\right\Vert \right.\right)\]
appeared at several places. The measures $\nu$ and $\mu^{*}\cdot\left\Vert \nu\right\Vert /\left\Vert \mu^{*}\right\Vert $
have the same norm and the divergence can also be written as $\left\Vert \nu\right\Vert \cdot D\left(\nu/\left\Vert \nu\right\Vert \left\Vert \mu^{*}/\left\Vert \mu^{*}\right\Vert \right.\right)$
so it essentially just a scaled version of KL-divergence. Information
divergence satisfies the following chain rule 
\[
D\left(\nu\Vert\mu\right)=D\left(\left\Vert \nu\right\Vert \Vert\left\Vert \mu\right\Vert \right)+D\left(\nu\left\Vert \mu^{*}\cdot\left\Vert \nu\right\Vert /\left\Vert \mu^{*}\right\Vert \right.\right).
\]
The first part $D\left(\left\Vert \nu\right\Vert \Vert\left\Vert \mu\right\Vert \right)$
is related to the uncertainty in the sample size, and the second part
$D\left(\nu\left\Vert \mu^{*}\cdot\left\Vert \nu\right\Vert /\left\Vert \mu^{*}\right\Vert \right.\right)$
is related to the uncertainty of the letters given the sample size.
If the sample size is fixed, we can use KL-divergence, but if the
sample size is not fixed and we have to code both the letters and
the sample size then have have to use information divergence rather
than KL-divergence. 

It was proved in \cite[Thm. 3.6]{Last2017}, that for any $s$-finite
measure $\mu$ on $\mathbb{A}$ there exists a point process $\omega\to\mu_{\omega}$
such that
\begin{itemize}
\item For any measurable set $A\subseteq\mathbb{A}$ the map $\omega\to\mu_{\omega}\left(A\right)$
is a Poisson distributed random variable with mean $\mu\left(A\right)$.
\item For any subsets $A_{1},A_{2}\subseteq\mathbb{A}$ the random variables
$\mu_{\omega}\left(A_{1}\right)$ and $\mu_{\omega}\left(A_{2}\right)$
are conditionally independent given $\mu_{\omega}\left(A_{1}\cap A_{2}\right)$.
\end{itemize}
This point process is called the \emph{Poisson point process} with
expectation measure $\mu$, and it will be denoted $Po\left(\mu\right).$
We conjecture that Poisson point processes also exists for s-finite
valuations.

For finite measures $\mu,\nu$ is is easy to prove that
\begin{equation}
D\left(\mu\Vert\nu\right)=D\left(\left.Po\left(\mu\right)\right\Vert Po\left(\nu\right)\right),\label{eq:PoissonInter}
\end{equation}
and the significance of this identity was pointed out in \cite{Harremoes2023,Lardy2024,Harremoes2025a}.
It is a sufficiency result inthe sense of Kullback and Leibler \cite{Kullback1951}.
Following \cite{Harremoes2025a} we will call this the \emph{Poisson
interpretation} of the measures $\mu$ and $\nu$. The goal is to
translate statements about the general measures $\mu$ and $\nu$
into statements about Poisson point processes. 
\begin{thm}
Let $\mu$ and $\nu$ denote the expectation measures of the Poisson
processes $Po\left(\mu\right)$ and $Po\left(\nu\right).$ If $D\left(\mu\Vert\nu\right)<\infty$
then Equation \eqref{eq:PoissonInter} holds.
\end{thm}
\begin{IEEEproof}
Using the decomposition in Proposition \ref{prop:Decomp} we get
\begin{multline*}
D\left(\mu\Vert\nu\right)=\\
D\left(\mu\left(\cdot\cap B\right)+\mu\left(\cdot\cap\complement B\right)\left\Vert \nu\left(\cdot\cap B\right)+\nu\left(\cdot\cap\complement B\right)\right.\right)\\
=D\left(\mu\left(\cdot\cap B\right)\left\Vert \nu\left(\cdot\cap B\right)\right.\right).
\end{multline*}
The result was recently proved for $\sigma$-finite measures in \cite[Thm. 5]{Leskelae2024},
so we get
\begin{align*}
D\left(\mu\Vert\nu\right) & =D\left(Po\left(\mu\left(\cdot\cap B\right)\right)\left\Vert Po\left(\nu\left(\cdot\cap B\right)\right)\right.\right)\\
 & =D\left(Po\left(\mu\right)\left\Vert Po\left(\nu\right)\right.\right).
\end{align*}
\end{IEEEproof}

\subsection{Information projection}

Let $\nu$ be a measure on $\mathbb{A}.$ Let $C$ denote a convex
set of measures on $\mathbb{A}.$ Then we define $D\left(C\Vert\nu\right)=\inf_{\mu\in C}D\left(\mu\Vert\nu\right).$
We say that $\nu^{*}$ is the information projection of $\nu$ on
$C$ if the following inequality is satisfied
\[
D\left(\mu\Vert\nu\right)\geq D\left(\mu\Vert\nu^{*}\right)+D\left(C\Vert\nu\right)
\]
for all $\mu\in C.$ 
\begin{lem}
Assume that $P$ is a point process with expectation measure $\mu$.
Then 
\[
D\left(P\Vert Po\left(\nu\right)\right)=D\left(P\Vert Po\left(\mu\right)\right)+D\left(Po\left(\mu\right)\Vert Po\left(\nu\right)\right).
\]
\end{lem}
\begin{prop}
Assume that $C$ is a convex set of measures and assume that $\nu$
is a measure for which there exists Poisson process $Po\left(\nu\right)$
with expectation measure $\nu.$ Let $\tilde{C}$ denote the set of
point processes on $\mathbb{A}$ with expectation measures belonging
to $C.$ Then
\[
D\left(\tilde{C}\Vert Po\left(\nu\right)\right)=D\left(C\Vert\nu\right).
\]
\end{prop}
\begin{IEEEproof}
Assume that a point process $P$ has expectation measure $\mu\in C$.
Then 
\begin{multline*}
D\left(P\Vert Po\left(\nu\right)\right)=D\left(P\Vert Po\left(\mu\right)\right)+D\left(Po\left(\mu\right)\Vert Po\left(\nu\right)\right)\\
\geq D\left(Po\left(\mu\right)\Vert Po\left(\nu\right)\right)=D\left(\mu\Vert\nu\right)\geq D\left(C\Vert\nu\right).
\end{multline*}
There the minimum over all point processes $P$ with expectation measures
in C equals the minimum over all point processes with expectation
measures in $C.$
\end{IEEEproof}
\begin{thm}
If $\nu$ is a measure and $\nu^{*}$ the information projection of
$\nu$ on the convex set $C$ then $Po\left(\nu^{*}\right)$ is the
information projection of $Po\left(\nu\right)$ on $\tilde{C}.$
\end{thm}
\begin{IEEEproof}
Assume that $P$ is a point process with expectation measure $\mu\in C.$
Then 
\begin{multline*}
D\left(P\Vert Po\left(\nu\right)\right)=D\left(P\Vert Po\left(\mu\right)\right)+D\left(Po\left(\mu\right)\Vert Po\left(\nu\right)\right)\\
=D\left(P\Vert Po\left(\mu\right)\right)+D\left(\mu\Vert\nu\right)\\
\geq D\left(P\Vert Po\left(\mu\right)\right)+D\left(\mu\Vert\nu^{*}\right)+D\left(C\Vert\nu\right)\\
=D\left(P\Vert Po\left(\mu\right)\right)+D\left(Po\left(\mu\right)\Vert Po\left(\nu^{*}\right)\right)+D\left(\tilde{C}\Vert Po\left(\nu\right)\right)\\
=D\left(P\Vert Po\left(\nu^{*}\right)\right)+D\left(\tilde{C}\Vert Po\left(\nu\right)\right).
\end{multline*}
\end{IEEEproof}

\subsection{Reverse information projection}

Let $C$ denote a convex set of measures and let $P$ denote a measures.
If $Q=\hat{P}\in C$ minimizes $D\left(P\Vert Q\right)$ over $Q\in C$
then $\hat{P}$ is called the reverse information projection of $P$
on $C.$ The notion of reverse information projections generalizes
the notion of maximum likelihood estimates. It has found applications
in safe testing and it has also been extended to cases where no measure
$Q\in C$ minimizes $D\left(P\Vert Q\right)$. 
\begin{thm}
Let $C$ denote a convex set of finite measures and let $\mu$ denote
a finite measure with reverse information projection $\hat{\nu}\in C.$
Then $Po\left(\hat{\nu}\right)$ is the reverse information projection
of Po$\left(\mu\right)$ on the convex hull of the measures of the
form $Po\left(\nu\right)$ where $\nu\in C.$
\end{thm}
\begin{IEEEproof}[Proof sketch]

Step 1: First assume that $\frac{\mathrm{d}P}{\mathrm{d}\hat{P}}$
only takes countably many different values. In this case the result
follows from \cite[Prop. 5]{Harremoes2025a}. Actually, the result
in \cite[Prop. 5]{Harremoes2025a} was formulated situations when
$\frac{\mathrm{d}P}{\mathrm{d}\hat{P}}$ only takes finitely many
different values, but the proof also works $\frac{\mathrm{d}P}{\mathrm{d}\hat{P}}$
when takes countably many different values.

Step 2: Write $\mathbb{A}$ as a countable disjoint union $\mathbb{A}=\bigcup A_{i}$
such that $\left(1+\epsilon\right)^{i}\leq\frac{\mathrm{d}P}{\mathrm{d}\hat{P}}\leq\left(1+\epsilon\right)^{i+1}$
on $A_{i}$ for $i\in\mathbb{Z}.$ Prove that 
\[
\int\frac{\mathrm{d}Po\left(\mu\right)}{\mathrm{d}Po\left(\hat{\nu}\right)}\,\mathrm{d}Po\left(\nu\right)\leq\exp\left(\epsilon\cdot\left\Vert \nu\right\Vert \right).
\]
Since this inequality holds for all $\epsilon>0,$ we have 
\begin{equation}
\int\frac{\mathrm{d}Po\left(\mu\right)}{\mathrm{d}Po\left(\hat{\nu}\right)}\,\mathrm{d}Po\left(\nu\right)\leq1\label{eq:Evariabel}
\end{equation}
for all $\nu\in C.$

Step 3: Since \eqref{eq:Evariabel} holds for all $\nu\in C$, we
have 
\begin{equation}
\int\frac{\mathrm{d}Po\left(\mu\right)}{\mathrm{d}Po\left(\hat{\nu}\right)}\,\mathrm{d}Q\leq1\label{eq:konvEvariabel}
\end{equation}
for any $Q$ in the convex hull of the Poisson distributions $Po\left(\nu\right),\,\nu\in C.$
Since \eqref{eq:konvEvariabel} holds for any $Q$ in the convex hull
of the Poisson distributions, the probability measure$Po\left(\hat{\nu}\right)$
is the reversed information projection of $Po\left(\mu\right)$ on
the convex hull of the Poisson distributions (see \cite[Lem. 4.1]{Li1999a}
and \cite[text after Thm. 5]{Lardy2024}). 
\end{IEEEproof}

\section*{Acknowledgment}

I would like to thank Tyron Lardy, Peter Grünwald, and Lasse Leskelä
for stimulating discussions, that have influenced the content of this
paper. 

\pagebreak{}




\end{document}